\documentclass[sigconf,screen,nonacm]{acmart}
\usepackage{microtype}
\usepackage{graphicx}
\usepackage{booktabs}
\usepackage{array}
\usepackage{xcolor}
\usepackage{float}
\graphicspath{{figures/}}
\newcommand{\sys}{\textsc{CTRL}}

\newif\ifprovisional\provisionalfalse
\newcommand{\provisional}[1]{\ifprovisional\textcolor{blue}{#1}\else#1\fi}
\newcommand{\pendingnote}[1]{\ifprovisional\textcolor{red}{[#1]}\fi}
\newcommand{\StalenessBound}{2}

\newcommand{\IntroStateGiB}{220}
\newcommand{\IntroShardGiB}{55}

\newcommand{\IntroEightBShardGiB}{30.5}

\newcommand{\IntroRCapGiB}{24}
\newcommand{\IntroRPeakGiB}{21.2}
\newcommand{\BgRResidentGiB}{38.1}
\newcommand{\BgRResidentOverGiB}{14.1}

\newcommand{\IntroQualityCtrl}{\provisional{79.4}}

\newcommand{\IntroQualityAsync}{\provisional{79.2}}

\newcommand{\BgMeasWindowGiB}{\provisional{5.4}}

\newcommand{\BgMeasKVGiB}{\provisional{38.1}}

\newcommand{\BgMicroTokK}{32}

\newcommand{\DesignWindowUnits}{3}

\newcommand{\EvQualQuestions}{500}
\newcommand{\EvQualUpdates}{150}

\newcommand{\LawKVInfeasible}{0.57}
\newcommand{\LawKVNeeded}{0.60}
\newcommand{\LawKVH}{47.8}

\newcommand{\HostImageGiB}{27.5}

\newcommand{\BgHMeanTok}{8500}
\newcommand{\BgHCapPct}{13.7}
\newcommand{\BgHGenPct}{53}
\newcommand{\BgHLearnerPct}{43}
\newcommand{\BgHSwitchPct}{4}

\newcommand{\BgABaseSwitchPct}{13}

\newcommand{\BgHMbppMeanTok}{2288}
\newcommand{\BgHMbppMedianTok}{1513}
\newcommand{\BgHMbppPNinetyTok}{5925}
\newcommand{\BgHMbppCapPct}{6.6}
\newcommand{\BgHMbppUpdateK}{658}
\newcommand{\BgHMbppLearnerPct}{40}
\newcommand{\BgHMbppGenPct}{52}
\newcommand{\BgHMbppSwitchPct}{9}

\newcommand{\AbRNoGateTok}{1592}
\newcommand{\AbRNoReuseTok}{1997}
\newcommand{\AbRNoGateX}{0.74}
\newcommand{\AbRNoReuseX}{0.93}

\newcommand{\ProfileCtrlActive}{63.1}
\newcommand{\ProfileColoActive}{46.1}
\newcommand{\ProfileCtrlIssue}{8.4}
\newcommand{\ProfileColoIssue}{3.2}
\newcommand{\ProfileCtrlTensor}{16.0}
\newcommand{\ProfileColoTensor}{11.2}

\newcommand{\NTZeroSwitchS}{0.6}
\newcommand{\NTSyncSwitchS}{14.1}

\input{tables/numbers-final}
\newcommand{\SLNTCo}{7468}
\newcommand{\SLNTSplit}{6759}
\newcommand{\SLNTTs}{5941}
\newcommand{\SLNTSplitX}{1.10}
\newcommand{\SLNTTsX}{1.26}
\newcommand{\SLTKCo}{5129}
\newcommand{\SLTKSplit}{3938}
\newcommand{\SLTKTs}{4656}
\newcommand{\SLTKSplitX}{1.30}
\newcommand{\SLTKTsX}{1.10}
\newcommand{\SLATCo}{3018}
\newcommand{\SLATSplit}{2148}
\newcommand{\SLATTs}{2593}
\newcommand{\SLATSplitX}{1.41}
\newcommand{\SLATTsX}{1.16}
\newcommand{\SLABCo}{3136}
\newcommand{\SLABSplit}{3164}
\newcommand{\SLABTs}{2549}
\newcommand{\SLABSplitX}{0.99}
\newcommand{\SLABTsX}{1.23}

\newcommand{\SLRMTsX}{1.43}

\newcommand{\SLRMLTsX}{1.34}
\newcommand{\SLNLargeX}{1.18}
\newcommand{\SLTsLo}{1.10}
\newcommand{\SLTsHi}{1.26}
\newcommand{\SLSplitLo}{1.10}
\newcommand{\SLSplitHi}{1.41}
\newcommand{\LedNTCo}{0.54}
\newcommand{\LedNTSplit}{0.59}
\newcommand{\LedNTTs}{0.67}
\newcommand{\LedNTHidden}{0.25}
\newcommand{\LedNTSlow}{0.13}
\newcommand{\LedTKCo}{0.78}
\newcommand{\LedTKSplit}{1.02}
\newcommand{\LedTKTs}{0.86}
\newcommand{\ResSplitActive}{83}  %
\newcommand{\ResCoActive}{96}  %

\newcommand{\AbMpsOffX}{0.87}  %
\newcommand{\SpanRows}{%
    H100 14B-NT, $N{=}256$ & 62.0 & 57.7 & 53.5 \\
    H100 14B-T MBPP, $N{=}256$ & 99.0 & 80.6 & 91.1 \\
    H100 14B-T 16K, $N{=}256$ & 438.4 & 365.5 & 364.8 \\
    A800 8B-T 16K, $N{=}128$ & 463.1 & 354.4 & 393.0 \\
    RTX~4090 8B-Base, $N{=}256$ & 98.2 & 73.5 & 78.8 \\}
\newcommand{\ModelPickAll}{25}
\newcommand{\ModelPickOk}{23}
\newcommand{\ModelPickLoss}{0.7}

\newcommand{\EightCtrlTok}{13814}

\newcommand{\EightVerlX}{1.23}

\newcommand{\EightSixSplitX}{1.42}

\newcommand{\AbsEightX}{\EightVerlX}

\newcommand{\AbsSplitHi}{1.30}
\newcommand{\StZeroCtrl}{5129}  %
\newcommand{\StTwoCtrl}{7303}  %
\newcommand{\StTwoCtrlReal}{2.00}  %
\newcommand{\StFourCtrl}{7788}  %
\newcommand{\StFourCtrlReal}{4.00}  %
\newcommand{\StTwoVerl}{6157}  %
\newcommand{\StFourVerl}{5898}  %
\newcommand{\StTwoVerlReal}{1.79}  %
\newcommand{\StFourVerlReal}{3.84}  %
\newcommand{\BlHState}{optimizer on host}
\newcommand{\BlHThinkState}{full host offload}
\newcommand{\BlAState}{optimizer on host}
\newcommand{\BlRState}{full host offload}

\newcommand{\HwHRollSmAct}{53}
\newcommand{\HwHRollIssue}{8}
\newcommand{\HwHRollTensor}{10}

\newcommand{\HwHLearnTensor}{56}
\newcommand{\HwHLearnDram}{26}

\newcommand{\HwHCoTensor}{51}

\newcommand{\HwRRollTensor}{4}

\newcommand{\HwRCoTensor}{22}

\newcommand{\KeepHRollSmall}{0.97} %

\newcommand{\KeepRSumLo}{1.24}     %
\newcommand{\KeepRSumHi}{1.31}
\newcommand{\KeepRPairLo}{1.05}    %
\newcommand{\KeepRPairHi}{1.08}
\newcommand{\KeepCapGain}{1.6}     %
\newcommand{\HwLearnShareCost}{12} %
\newcommand{\LawKVHAnchor}{44.6}
\renewcommand{\LawKVInfeasible}{0.54}

\newcommand{\OlyQuestions}{536}
\newcommand{\OlyCtrlAcc}{63.6}  %
\newcommand{\OlyColoAcc}{63.3}  %

\newcommand{\StZeroVerl}{5159}
\renewcommand{\NTZeroSwitchS}{1.2}  %
\renewcommand{\NTSyncSwitchS}{5.8}
\renewcommand{\BgHMeanTok}{8516}     %
\renewcommand{\BgHCapPct}{13.8}
\renewcommand{\BgHGenPct}{55}
\renewcommand{\BgHLearnerPct}{44}
\renewcommand{\BgHSwitchPct}{1}
\renewcommand{\BgABaseSwitchPct}{10} %
\renewcommand{\BgHMbppMeanTok}{2380} %
\renewcommand{\BgHMbppMedianTok}{1505}
\renewcommand{\BgHMbppPNinetyTok}{6354}
\renewcommand{\BgHMbppCapPct}{7.1}
\renewcommand{\BgHMbppUpdateK}{678}  %
\renewcommand{\BgHMbppLearnerPct}{37}
\renewcommand{\BgHMbppGenPct}{59}
\renewcommand{\BgHMbppSwitchPct}{4}

\renewcommand{\IntroRCtrlTok}{2076}
\renewcommand{\IntroRColocRatio}{1.42}
\renewcommand{\IntroRSyncGenPct}{56}
\renewcommand{\IntroRSyncLearnerPct}{39}
\renewcommand{\IntroRSyncSwitchPct}{5}
\renewcommand{\ProfileCtrlActive}{57.7}
\renewcommand{\ProfileColoActive}{47.1}
\renewcommand{\ProfileCtrlIssue}{3.9}
\renewcommand{\ProfileColoIssue}{3.2}
\renewcommand{\ProfileCtrlTensor}{15.0}
\renewcommand{\ProfileColoTensor}{10.0}

\newcommand{\AbRCtrlTok}{2076}
\renewcommand{\AbRNoGateTok}{1835}
\renewcommand{\AbRNoGateX}{0.88}
\renewcommand{\AbRNoReuseTok}{1955}
\renewcommand{\AbRNoReuseX}{0.94}
\newcommand{\AbTsX}{0.83}   %

\newcommand{\MbppLo}{1.71}   %
\newcommand{\MbppHi}{2.24}  %
\newcommand{\MbppActorLo}{14}
\newcommand{\MbppActorHi}{18}

\newcommand{\MbppStepLo}{27}  %
\newcommand{\MbppStepHi}{43}
\newcommand{\MbppCtrlPubS}{0.3}
\newcommand{\MbppCtrlLo}{12}
\newcommand{\MbppCtrlHi}{20}

\newcommand{\SweepRLearnPct}{8}
\newcommand{\SweepRSpanPct}{13}
\renewcommand{\IntroQualityCtrl}{79.4}
\renewcommand{\IntroQualityAsync}{79.2}

\newcommand{\OlyColoTokX}{1.27}

\newcommand{\StFourOverTwoPct}{7}  %

\newcommand{\HwRRollSmAct}{35}
\renewcommand{\HwRRollTensor}{3}

\renewcommand{\HwRCoTensor}{20}

\newcommand{\MbppSwitchS}{10}  %
\renewcommand{\DesignWindowUnits}{3}
\renewcommand{\BgMeasKVGiB}{44.7}
\renewcommand{\IntroRPeakGiB}{23.5}

\newcommand{\draftnote}[1]{}
\renewcommand{\pendingnote}[1]{\textcolor{red}{#1}}

\newcommand{\ctrlinstitution}{The Hong Kong University of Science and Technology (Guangzhou)}
\newcommand{\hkustgz}{%
  \affiliation{%
    \institution{\ctrlinstitution}
    \city{Guangzhou}
    \country{China}
  }%
}

\makeatletter
\newcommand{\ctrl@sharedaffiliation}[2]{\let\and\@typeset@author@bx}
\patchcmd{\@mkauthors@iii}
  {\leavevmode\lineskip=1pc\relax\centering\hspace*{-1em}%
   \addresses\let\and\@typeset@author@bx\and\par}
  {\centering\noindent\hbox to \textwidth{%
     \hfil
     \setlength{\author@bx@wd}{.29\textwidth}%
     \setlength{\author@bx@sep}{.5em}%
     \let\@authorfont\Large
     \hypersetup{urlcolor=black}%
     \let\affiliation\ctrl@sharedaffiliation
     \addresses\let\and\@typeset@author@bx\and\unskip
     \hfil
   }\par
   \smallskip{\normalsize\ctrlinstitution\par}}
  {}{\PackageError{ctrl-arxiv}{Could not format the shared author affiliation}{}}
\makeatother

\begin{document}
\title[CTRL: Concurrent Training and Rollout]{CTRL: Concurrent Training and Rollout for LLM Reinforcement Learning via GPU Co-Residency}
\author{Ruijia Yang}
\email{ryang379@connect.hkust-gz.edu.cn}
\hkustgz
\author{Ruikai Li}
\email{rli321@connect.hkust-gz.edu.cn}
\hkustgz
\author{Zeyi Wen}
\email{wenzeyi@hkust-gz.edu.cn}
\hkustgz
\renewcommand{\shortauthors}{Yang, Li, and Wen}

\begin{abstract}
Reinforcement learning (RL) post-training of large language models runs
two GPU workloads with different hardware demands: rollout decoding,
which is dominated by memory access and whose batch shrinks as responses
finish, and a dense learner. Current systems either time-share every GPU
between them, paying serial phases and switches, or split GPUs into
pools whose sizes change only in whole devices. We present \sys, which
makes co-residency a third RL execution layout: every GPU runs a learner
rank beside a persistent rollout engine, sharing the device under CUDA
MPS. A host-owned learner makes the layout fit: optimizer state stays
in host memory, and policy weights stream layer by layer through a bounded
GPU window. A publication gate trains each ready batch first and waits
only before releasing the next policy, keeping a dropless staleness
bound. A step-time bound guides compute-share selection, and a GPU-time
ledger explains overlap and sharing costs. On H100, A800,
and RTX~4090 GPUs with Qwen3 models, \sys\ achieves \HeadBestLo--\HeadBestHi$\times$
the throughput of the stronger of verl's asynchronous baselines across
\HeadRows\ workload settings with the staleness limit set to two, and \AbsEightX$\times$
verl's \texttt{colocate\_async} on eight H100 GPUs. With the same learner, co-residency is
\SLTsLo--\SLTsHi$\times$ faster than time-sharing on all three platforms and
1.10--\AbsSplitHi$\times$ faster than the best measured whole-GPU split on H100
DAPO-Math workloads at 256--528 trajectories per step, on four and eight
GPUs.

\end{abstract}
\maketitle

\section{Introduction}
\label{sec:introduction}
\begin{figure}[!t]
  \centering
  \includegraphics[width=\columnwidth]{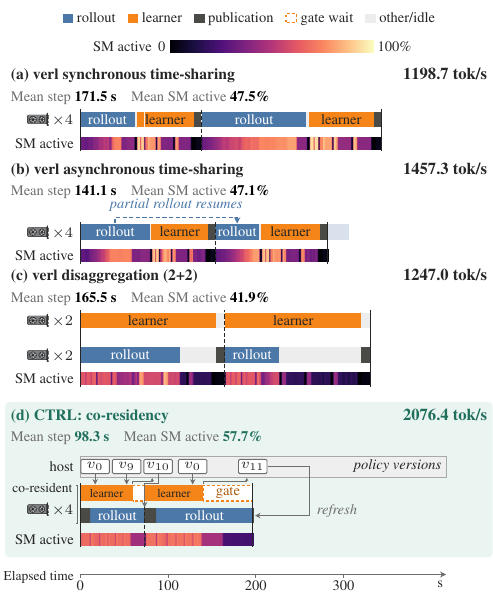}
  \caption{Execution layouts in the running example over two
  steady-state steps, drawn to each system's mean step time, with SM
  activity beneath. Host slot $v_0$ is the reference policy.}
  \ifdefined\Description
  \Description{Four systems in the original single-column arrangement.
  Time-sharing uses one execution row; disaggregation and CTRL use
  learner and rollout rows, with SM active beneath. The figure covers
  consecutive captured steps 10 and 11, scaled jointly to twice the
  formal mean step time, including the between-step gap. Orange denotes reference
  scoring and actor training, blue denotes rollout host requests,
  gray overlays publication, and dashed orange boxes denote gate wait.
  The CTRL host strip illustrates reference and policy-version flow;
  publication can cross step boundaries. All systems share the same
  mapped-seconds scale and four-GPU SM color scale. Heatstrips use two-second
  triangular smoothing in capture time; printed SM means integrate unsmoothed
  four-GPU measurements over the continuous window from step 5 entry to
  step 11 exit, including between-step gaps, and divide by its duration. Printed mean step
  uses the untraced updates 5--12 continuous duration divided by eight,
  averaged within seeds and then across seeds.}
  \fi
  \label{fig:overview-modes}
\end{figure}

Reinforcement learning (RL) post-training of reasoning
models~\citep{deepseekmath,dapo} runs two GPU workloads that stress the
hardware differently. \emph{Rollout} generates groups of responses by
autoregressive decoding: every step rereads the weights and KV state, so
memory access, not computation, dominates it, and its active batch
shrinks as responses finish. The \emph{learner} evaluates reference-policy
probabilities and trains the actor on the generated tokens with dense
computation over complete sequences. Long reasoning responses increase
the work of both. How the two engines share a GPU allocation therefore
determines training throughput.

Existing RL systems share GPUs in time or partition them in space
(Figure~\ref{fig:overview-modes}a--c). \emph{Synchronous time-sharing}
gives every GPU to one phase at a time. In our running example,
Qwen3-8B-Base~\citep{qwen3} on DAPO-Math with four RTX~4090s,
verl's hybrid engine~\citep{hybridflow} spends \IntroRSyncGenPct\% of a
step generating, \IntroRSyncLearnerPct\% in the learner, and
\IntroRSyncSwitchPct\% switching weights; only one engine makes progress
at a time, and every switch sits on the critical path.
\emph{Asynchronous time-sharing}, as in verl's \texttt{colocate\_async},
adds partial rollout: unfinished responses continue across policy
updates, but learner work still blocks generation. \emph{Disaggregation}
overlaps the phases on separate GPU pools~\citep{areal,streamrl,rollflash},
but splitting the GPUs shrinks each engine's share and sizes the pools
in whole devices: in the running example, verl's 2+2 split gives each
engine only two of the four GPUs.
Either way capacity is stranded: a time-shared GPU serves one engine at
a time, and a pool waits whenever the other pool is slower.

The two workloads are complementary on a single GPU. Alone on an
H100 node, rollout keeps the tensor pipes active \HwHRollTensor\% of
the time and issues instructions in \HwHRollIssue\% of SM cycles; the
learner keeps the tensor pipes active \HwHLearnTensor\% of the time and
uses \HwHLearnDram\% of the memory bandwidth. Running a learner rank beside a rollout
shard on every GPU lets both engines progress on each device, puts the
computation that decoding leaves unused to work, and removes phase
switching.
The obstacle is memory: in the running
example, a four-way shard of GPU-resident training state alone occupies
\IntroEightBShardGiB\,GiB per GPU, more than the device's
\IntroRCapGiB\,GiB, before rollout weights, activations, or KV cache.

Time-sharing systems can already offload training state to host
memory, yet they suspend rollout and release its KV cache during learner
execution. Recent layer-streaming
trainers~\citep{zero_offload,zero_infinity,slideformer,slidedp} point to
an opportunity: they keep training state on the host and stream layers
through a bounded GPU window, so a learner needs only that window, not
the whole device. Our key insight is to \emph{organize the RL policy
lifecycle around host-owned state and a bounded GPU execution window},
making room for a persistent rollout engine on every GPU. \sys\ extends this bounded footprint across the RL policy lifecycle:
immutable host versions support reference scoring, actor training, and
policy delivery while rollout stays resident. Incremental policy export prepares the next version in the background,
while reuse of resident rollout weights reduces host transfers. Learner and
rollout together peak at \IntroRPeakGiB\,GiB per GPU in the running
example.

The result is a third RL execution layout beside time-sharing and
disaggregation: every GPU hosts a data-parallel learner rank beside a
rollout shard. CUDA Multi-Process Service (MPS)~\citep{nvidia_mps} lets
their kernels share the device; the policy lifecycle makes this sharing
possible throughout scoring, training, and refresh.

Synchronization then determines how much work overlaps: a batch can be
ready to train while unfinished rollout still delays publication of the
next policy. Host-owned versions separate the two events: the learner
commits a new policy to the host while rollout keeps decoding with the
current one. \sys\ therefore trains the ready batch first and places
the necessary wait at a \emph{publication gate} before the new policy
reaches rollout, preserving the staleness bound without dropping
responses. Unlike Rollplex~\citep{rollplex}, which
overlaps response-independent prefix work with decoding, \sys\ overlaps
complete learner updates with rollout for subsequent batches
(Figure~\ref{fig:overview-modes}d; \S\ref{sec:related}).

Co-residency also makes the engines share compute and memory
bandwidth. \sys\ uses a step-time bound based on learner duration and
rollout span to guide probes of the learner's compute share within a
feasible memory budget. Measured throughput determines whether to
adopt a new share.

Our contributions follow the three requirements of co-residency:
fitting, overlapping, and balancing the engines.

\textbf{1. A host-owned learner for RL.} Layer streaming moves from
fitting one model to serving the RL policy lifecycle: reference scoring,
actor training, and policy delivery share immutable host versions and
one bounded GPU window beside a persistent rollout engine, with
incremental export, shard-directed refresh, and reuse of rollout
weights.

\textbf{2. A co-resident layout with a publication gate.} Complete
learner updates overlap rollout for later batches under a dropless
staleness bound; persistent rollout state removes phase switching, and
the publication gate places the necessary wait after training.

\textbf{3. Configuration guided by resource balance.} We derive a
step-time bound and use it to guide short measurements of the learner's
compute share within a feasible memory budget. Retrospective share
sweeps across three platforms evaluate the resulting ranking of
configurations. A GPU-time ledger explains each layout's overlap and
exposed costs.

We implement \sys\ on verl and vLLM~\citep{hybridflow,vllm} and evaluate
Qwen3 Base, thinking, and non-thinking models on math and code workloads
across four-GPU H100, A800, and RTX~4090 nodes. With a common configured
staleness limit of \StalenessBound, \sys\ delivers
\HeadBestLo--\HeadBestHi$\times$ the throughput of the stronger
asynchronous baseline across all workloads.
The running example reaches \IntroRCtrlTok\ tokens/s,
\IntroRColocRatio$\times$ asynchronous time-sharing, with
\ProfileCtrlActive\% SM activity versus \ProfileColoActive\%
(Figure~\ref{fig:overview-modes}). With the same learner, co-residency is
\SLTsLo--\SLTsHi$\times$ faster than time-sharing and up to
\SLSplitHi$\times$ faster than the best measured whole-GPU split.
The GPU-time ledger attributes the gain to rollout work overlapped
with learning. Hardware counters show the complementary demands:
on H100, co-residency raises tensor-pipe activity from
\HwHRollTensor\% to \HwHCoTensor\% with little change in memory traffic.

\section{Background and Motivation}
\label{sec:background}

\begin{table}[H]
  \caption{Workloads and synchronous phase shares.}
  \label{tab:workloads}
  \centering\scriptsize
  \setlength{\tabcolsep}{2pt}
  \begin{tabular}{@{}>{\raggedright\arraybackslash}p{0.24\columnwidth}rrrrrrc@{}}
    \toprule
     & & \multicolumn{3}{c}{Response length} & At cap & Tok. & L/G/S \\
    \cmidrule(lr){3-5}
    Workload & $N$ & mean & p50 & p90 & (\%) & (k) & (\%) \\
    \midrule
    4090 8B-B DAPO-8K$^\ast$ & 256 & 656 & 527 & 1067 & 0.1 & 206 & \IntroRSyncLearnerPct/\IntroRSyncGenPct/\IntroRSyncSwitchPct \\
    A800 8B-B DAPO-8K & 128 & 658 & 512 & 1063 & 0.3 & 102 & 19/71/10 \\
    H100 14B-NT DAPO-8K & 256 & 1600 & 1428 & 2884 & 0.1 & 452 & 42/52/7 \\
    A800 8B-T DAPO-16K & 128 & 9405 & 8519 & 16384 & 19.1 & 1227 & 28/70/1 \\
    H100 14B-T DAPO-16K & 256 & \BgHMeanTok & 7498 & 16384 & \BgHCapPct & 2242 & \BgHLearnerPct/\BgHGenPct/\BgHSwitchPct \\
    \midrule
    4090 8B-B MBPP-8K & 256 & 70 & 47 & 136 & 0.0 & 55 & 47/32/21 \\
    A800 8B-T MBPP-8K & 256 & 2980 & 2047 & 8192 & 10.3 & 807 & 34/63/3 \\
    H100 14B-T MBPP-8K & 256 & \BgHMbppMeanTok & \BgHMbppMedianTok & \BgHMbppPNinetyTok & \BgHMbppCapPct & \BgHMbppUpdateK & \BgHMbppLearnerPct/\BgHMbppGenPct/\BgHMbppSwitchPct \\
    \bottomrule
  \end{tabular}\\[3pt]
  \parbox{\columnwidth}{\scriptsize
  Qwen3 models; B: Base; T/NT: thinking/non-thinking.
  8K/16K: response caps; At cap: fraction reaching the cap.
  Tok.: prompt + response tokens per step (k).
  L/G/S: learner (reference + actor), generation, switching (rounded \%).
  $^\ast$Running example.}
\end{table}

\subsection{RL workloads and policy dependencies}
\label{sec:bg-loop}

We consider GRPO-style RL post-training~\citep{ppo,deepseekmath,dapo}. Each training \emph{step} consumes $N=QG$ trajectories: $G$ responses to each of $Q$ prompts.
A verifiable reward scores the responses, and within-group normalization
produces their advantages. The learner evaluates reference-policy
log-probabilities, runs the actor's forward and backward passes, and updates the policy; \emph{update} refers only to this change of parameters, and the \emph{step time} is the wall-clock time of one step. A \emph{learner rank} processes its share of these
trajectories over the complete model. A \emph{rollout replica} generates
responses; when it spans GPUs, each holds a \emph{shard} of
its weights and KV cache.

Response lengths couple the two phases: total tokens determine learner
work, while unfinished responses determine when a rollout batch becomes
available. Response count alone does not set this balance
(Table~\ref{tab:workloads}): on A800 at the same $N=128$, thinking mode
supplies twelve times as many tokens per step as the Base model, and
thinking responses have a median of 8519 tokens, yet their 90th
percentile reaches the 16K cap, where 19.1\% are truncated. Resource
sharing must therefore account for both learner work and the pattern of
response completion.

The phases are also linked by policy versions: the learner consumes
generated responses and produces the policy for later generation.
Asynchronous execution allows responses to span policy versions, so training
can consume responses generated by older policies. A staleness bound
limits this version lag. Preserving responses requires waiting before further progress violates
the bound, making wait placement part of execution design (\S\ref{sec:design-staleness}).

\subsection{Execution layouts and their costs}
\label{sec:bg-layouts}

\paragraph{Synchronous time-sharing.} verl's hybrid
engine~\citep{hybridflow} gives each phase all GPUs in turn. FSDP2's CPU
offload can keep training state on the host, and rollout releases its KV
cache while the learner runs, but the engines still take turns on
every device. In the running example, the learner and the weight switch
take \IntroRSyncLearnerPct\% and \IntroRSyncSwitchPct\% of each step while
rollout waits (Figure~\ref{fig:overview-modes}a). Switching costs most in short synchronous steps: 21\% on RTX~4090 MBPP,
\BgABaseSwitchPct\% on A800 Base, 7\% on H100 non-thinking, and
\BgHSwitchPct\% on H100 thinking (Table~\ref{tab:workloads}).

\paragraph{Asynchronous time-sharing.} verl's \texttt{colocate\_async}
maintains partial rollouts across steps. Once a batch is available
under the staleness bound, it suspends unfinished requests, trains and
publishes the updated policy, and then resumes those responses, so a
ready batch trains without waiting for every in-flight response. Rollout still waits through learning and weight switching
(Figure~\ref{fig:overview-modes}b).

\begin{figure}[t]
  \centering
  \includegraphics[width=\columnwidth]{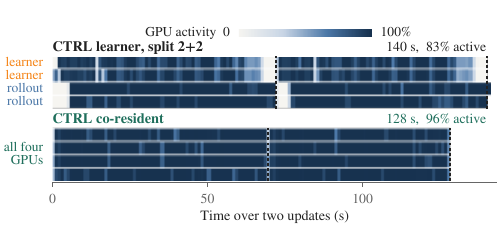}
  \caption{The same learner on four H100s (Qwen3-14B non-thinking,
  DAPO-Math) as a 2+2 split and co-resident in \sys: per-GPU activity
  over two training steps.}
  \label{fig:resource}
\end{figure}

\paragraph{Disaggregation.} Disaggregated systems overlap the phases on
separate GPU pools~\citep{areal,streamrl,rollflash}. They avoid phase
serialization, but each pool has fewer devices, the slower pool sets the
step time, and pool sizes change only in whole-GPU increments. In the
running example's 2+2 layout, the learner phase on two GPUs takes longer
than on four, and each pool idles while the other finishes
(Figure~\ref{fig:overview-modes}a,c).
Whole-GPU pools also strand capacity. With the same learner on four
H100s, a 2+2 split keeps its GPUs active \ResSplitActive\% of the time:
the rollout GPUs wait at the staleness bound while the learner GPUs
finish (Figure~\ref{fig:resource}). With both engines on every GPU, the
same hardware is active \ResCoActive\% of the time.

\subsection{Same-GPU co-residency: opportunity and requirements}
\label{sec:bg-sharing}

\begin{figure}[t]
  \centering
  \includegraphics[width=\columnwidth]{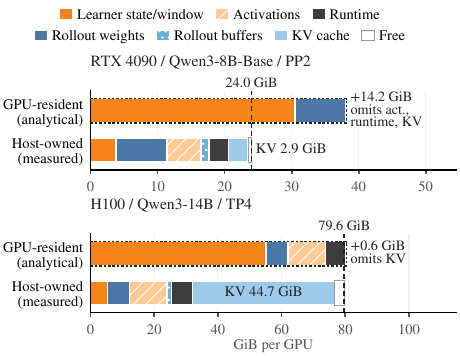}
  \caption{Per-GPU memory. Host-owned bars are measured at the
  whole-device peak; GPU-resident bars are analytical.}
  \label{fig:memory-budget}
\end{figure}

\begin{figure}[t]
  \centering
  \includegraphics[width=\columnwidth]{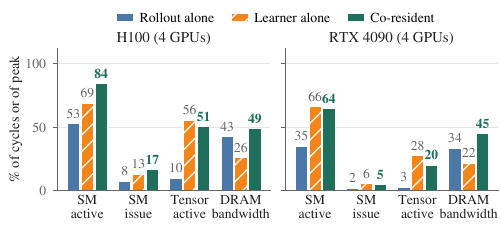}
  \caption{Device-wide Nsight Systems metrics during computation:
  engines alone and co-resident on four GPUs.}
  \label{fig:corun-counters}
\end{figure}

\begin{figure*}[t]
  \centering
  \includegraphics[width=\textwidth]{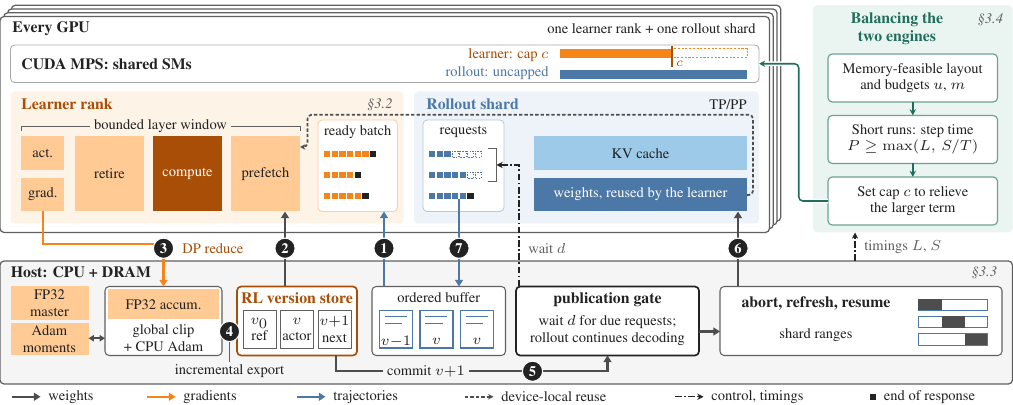}
  \caption{\sys's host-owned policy lifecycle and co-resident execution.}
  \label{fig:overview}
\end{figure*}

Co-residency combines the full device allocation of time-sharing with
the phase overlap of disaggregation, and CUDA MPS~\citep{nvidia_mps}
allows kernels from both processes to share each GPU. It pays because
the two engines load different parts of a GPU
(Figure~\ref{fig:corun-counters}).

\paragraph{What each engine leaves unused.} A decoding step is a few
hundred microsecond-scale kernels that mostly reread weights and KV
state. Alone on an H100 node, rollout holds warps on the SMs
\HwHRollSmAct\% of the time, yet the SMs issue instructions in only
\HwHRollIssue\% of cycles and the tensor pipes are active
\HwHRollTensor\% of the time. On RTX~4090, where most of a step decodes
the long tail of the batch, the SMs hold warps only \HwRRollSmAct\% of
the time and the tensor pipes are active \HwRRollTensor\%. The learner is the opposite: its tensor pipes
are active \HwHLearnTensor\% of the time on H100 while it uses
\HwHLearnDram\% of the memory bandwidth. When both run on the same
GPU, tensor activity rises from \HwHRollTensor\% to \HwHCoTensor\% on
H100 and from \HwRRollTensor\% to \HwRCoTensor\% on RTX~4090, while
memory traffic rises far less: the computation that decoding leaves
unused is put to work.

Co-residency requires a bounded footprint throughout the step and a
schedule that sustains overlap. A direct composition of verl's fully
offloaded FSDP2 learner and a resident vLLM engine under MPS illustrates
both requirements. On RTX~4090, rollout memory fractions up to 0.40
leave insufficient KV capacity for one maximum-length request; larger
fractions exhaust memory during the learner update or first publication.
On H100, the composition reaches 0.81$\times$ verl's time-sharing
throughput; \sys\ is 1.46$\times$ faster.

\paragraph{The memory barrier.} Both working sets must first fit.
Mixed-precision Adam state takes 16 bytes per parameter. In the running
example, the four-way shard of Qwen3-8B training state is
\IntroEightBShardGiB\,GiB; with the measured rollout weights it reaches
\BgRResidentGiB\,GiB, \BgRResidentOverGiB\,GiB beyond the RTX~4090
before any activations, runtime, or KV cache
(Figure~\ref{fig:memory-budget}). For Qwen3-14B on H100, the
\IntroShardGiB\,GiB shard with the same activations, rollout weights,
and runtime already exceeds capacity before any KV
cache. With host-owned state, both engines fit: the running example
peaks at \IntroRPeakGiB\,GiB per GPU, and on H100 a
\BgMeasWindowGiB\,GiB learner window leaves \BgMeasKVGiB\,GiB for KV
cache.

\paragraph{Policy lifecycle in a bounded footprint.} The time-sharing
baselines build a separate reference model and materialize updated
parameters on the GPU for transfer to rollout. A co-resident learner
must score, train, and deliver policies within a bounded
footprint beside resident rollout state.

\paragraph{Overlap under a version bound.} A ready training batch can
coexist with unfinished responses from other batches. Waiting for those
responses before training exposes their tail on the critical path. The schedule must overlap ready learner work with rollout while enforcing the version bound without dropping responses.

\paragraph{Balanced sharing.} Both engines compete for compute and
memory bandwidth within the same memory budget. Their different
responses to allocation require identifying which engine
sets the step time and relieving it without starving the other.
Sections~\ref{sec:design-window}--\ref{sec:design-config} address them
in turn.

\section{Design of \sys}
\label{sec:design}
\providecommand{\flow}[1]{{\usefont{U}{pzd}{m}{n}\char\numexpr181+#1\relax}}

\subsection{Overview}
\label{sec:design-overview}

\sys\ keeps a learner rank and a rollout shard on every GPU
(Figure~\ref{fig:overview}). A rollout replica spans one or more GPUs
through tensor or pipeline parallelism, so every GPU holds one rollout
shard; the learner ranks form one data-parallel group. The host holds the
optimizer state and committed policy versions. Each
learner rank operates on a bounded layer window beside the rollout
weights and KV cache.

Configuration first chooses a memory-feasible rollout layout for the
interconnect, then uses short measurements to balance the learner's
compute share against rollout (\S\ref{sec:design-coexec}).
In each step, the learner takes prompt groups in dispatch order once
complete, checking their behavior versions before training (\flow{1}). It
streams the reference and actor versions through its layer window
(\flow{2}), reduces gradients across ranks into the host accumulator
(\flow{3}), and runs CPU Adam while exporting each finished range into
the next policy image, which it then commits (\flow{4}). Rollout keeps
generating except during the brief refresh. The committed version waits at the publication
gate only while due requests remain in flight (\flow{5}); an
abort--refresh--resume transaction then loads each rollout shard's
ranges (\flow{6}). Completed responses return to the ordered buffer with
their behavior versions (\flow{7}). These timings guide configuration.

\subsection{A host-owned learner for RL}
\label{sec:design-window}

Reference scoring, actor training, and policy delivery must all fit
beside rollout. \sys\ extends host-resident layer streaming with three
RL lifecycle rules: readers hold immutable policy images, a new image
becomes visible only after its update completes, and rollout-weight
reuse ends before refresh overwrites its source. These rules let the
learner advance while rollout retains its current policy.

\paragraph{Shared versions for RL consumers.} Scoring uses a fixed
reference policy, training uses the current actor, and rollout receives
the next policy. \sys\ serves these consumers from immutable BF16
images in a shared host version ring.
A reader leases an image, preventing its slot from being overwritten.
The reference pass retains version~0; the actor reads the current
committed policy; publication leases the image being delivered to
rollout. With one publication per step, the ring has at least three
slots for the reference, current policy, and next commit. Readers take
what they need from host memory, so none gathers a full policy on a GPU.

\begin{figure}[t]
  \centering
  \includegraphics[width=\columnwidth]{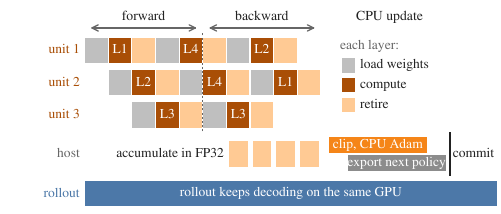}
  \caption{Layer-window execution through the final micro-batch and
  host update, beside persistent rollout (schematic).}
  \label{fig:window}
\end{figure}

\paragraph{One window for scoring and training.} Reference scoring and
actor training stream their versions through the same pool of $k$ GPU
layer units (Figure~\ref{fig:window}). Forward-only scoring releases gradient storage and returns
its units when the pass ends; training reuses the window for the PPO
forward and backward passes. A committed image is already BF16, so
filling a unit takes a copy rather than a conversion from FP32 master
weights. Within a pass, the layer executor applies established prefetch,
activation offload, and recomputation
techniques~\citep{zero_offload,zero_infinity,slideformer}; a unit
becomes reusable once its outstanding transfers and gradient operations
complete. Prefetching one layer ahead and reusing completed units makes
the pool size independent of model depth. For Qwen3-14B on H100, $k=\DesignWindowUnits$ units and a
\BgMicroTokK{}k-token micro-batch budget give a \BgMeasWindowGiB\,GiB
window. The 152K-vocabulary output layer is the largest: the actor
update evaluates it in chunks of tokens with verl's output-layer
function~\citep{hybridflow}, so the full logits are never materialized.

\paragraph{Incremental export during CPU updates.} Gradients are
combined across data-parallel ranks and accumulated in FP32 on the host
across micro-batches. Without GPU peer access, ranks reduce through
shared host memory: each copies its gradients there with the copy
engine, leaving the SMs to rollout, and the CPU sums them. One rank, the optimizer owner, applies CPU Adam to
the shared host master weights and moments. Exact global clipping needs
the final gradient of every layer, so the owner retains these gradients
until the last micro-batch completes, computes the norm, and then clips
and updates. As Adam finishes each parameter range, a background exporter
materializes that range in the next BF16 image, so version construction
overlaps the remaining CPU updates. The image is committed atomically
once all ranges are ready: readers never observe a partially updated
policy, and rollout continues throughout finalization. Publication then
leases the committed image and streams each shard's ranges
(\S\ref{sec:design-staleness}).

\paragraph{Rollout weights as a local source.} When a local rollout shard
holds the committed version that the learner needs, its weight tensors
supply the matching portion of a layer through CUDA IPC and a
device-local copy; the remaining portion comes from the host. For a TP
shard, the local slice and its host-supplied complement together fill
the layer. For PP, layers on the local stage come from rollout and the
other layers come from the host. Before
publication overwrites the rollout weights, the learner revokes this
source and waits for outstanding reads. Resident rollout weights
thus become a local source for the learner, reducing host traffic.

\subsection{Co-resident execution under bounded staleness}
\label{sec:design-staleness}

\paragraph{Sharing SMs with MPS\@.} Each GPU hosts one data-parallel
learner rank and one rollout shard as separate CUDA MPS
clients~\citep{nvidia_mps}. MPS admits kernels from both processes to
the shared SMs. The learner's active-thread cap, its compute share,
limits its access, while
rollout remains uncapped. The cap decides how the two engines divide a GPU. Their
kernels differ by orders of magnitude: a decoding step issues hundreds
of microsecond-scale kernels, and a learner layer a few
millisecond-scale kernels that span every SM, behind which decode
kernels queue. The cap keeps part of the SMs free of learner kernels,
so decode kernels wait less behind them: on H100, lowering
the learner's share from 99\% to 70\% raises the rate rollout keeps
beside it by \KeepCapGain$\times$. It reserves no exclusive SMs, so rollout can use
the resources released when the learner waits, and both engines keep
their GPU allocations.

The alternative to MPS is CUDA green contexts, which confine each
workload to a set of SMs inside one process. Without MPS, kernels from
different processes do not execute concurrently, so this route merges
the trainer and the inference engine into a single process, a
restructuring of both frameworks. MPS ships with the driver and keeps both engines in separate,
replaceable processes.

\paragraph{Ordered consumption and the version bound.} A cohort is a
group of prompts dispatched together, each with $G$ responses. The
controller consumes prompt groups in dispatch order, waiting for each
group to complete. Every trajectory
$y$ records its earliest behavior version $b(y)$, including when a
partial response resumes after a weight refresh. If step $g$ trains
policy $g-1$ on batch $B$ and produces policy $g$, its training staleness
is
\begin{equation}
  \sigma(B)=(g-1)-\min_{y\in B}b(y)\le T.
  \label{eq:staleness}
\end{equation}
The sampler validates this condition before each step. This consumption check uses actual behavior versions;
the publication gate separately governs requests still in flight.

\begin{figure}[t]
  \centering
  \includegraphics[width=\columnwidth]{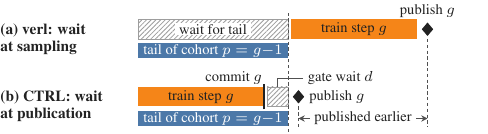}
  \caption{Publication-gated overlap at $T=2$ (schematic).}
  \label{fig:staleness}
\end{figure}

\paragraph{Train first, wait before publication.} Because each version
is committed to the host before rollout sees it (\S\ref{sec:design-window}),
training a version and publishing it are separate events. A complete
batch can be ready while a younger cohort is still decoding. verl's dropless sampler may wait for
that cohort before admitting the ready batch. \sys\ executes the ready learner update first and applies the dispatch-span wait immediately before
publication. For $T>0$, version $g$ waits for pending or running
prompts dispatched at step $p$ with
\begin{equation}
  g-p+1\ge T.
  \label{eq:publication-deadline}
\end{equation}
The gate opens when no pending or running request meets this condition.
During the wait, no new cohort is dispatched and rollout continues on
the published policy.

FIFO consumption closes the bound for completed responses as well.
With $W\le T$ warmup cohorts and one equally sized cohort dispatched and
consumed per update, cohort $k$ is dispatched at
$p(k)=\max(1,k-W)$ and consumed at update $k$. Admission resumes only
after the preceding policy is installed, giving every response
$b(y)\ge p(k)-1$, including responses retained across refreshes.
Thus $(k-1)-b(y)\le k-p(k)\le W\le T$: completion order cannot age a
buffered response beyond the bound. With eventual computation and
transfer completion, the next FIFO cohort becomes ready. Gate-blocking
requests complete on the current policy, so publication can proceed.

For example, at $T=2$, step $g$ trains its complete batch while the
cohort dispatched at $g-1$ finishes (Figure~\ref{fig:staleness}). If
that cohort completes during training, publication proceeds without a
tail wait; if it outlasts training, only the remaining tail delays
publication. Applying the same deadline before sampling would expose the
entire tail before learner computation begins. The gate thus turns the learner update into overlap within the
staleness bound.
At $T=0$ the gate drains in-flight rollout, serializing the engines
(\S\ref{sec:eval-time}).

\paragraph{Abort, refresh, and resume.} Once the gate opens, publication
pauses admission and aborts active decode requests, saving their partial
responses for resumption. The owner leases the committed BF16 image and
describes the byte ranges belonging to each rollout shard: TP slices or
PP layers. After outstanding learner reads of rollout weights finish,
each shard loads its ranges through pinned host staging buffers into its
existing weight tensors. All shard loads must be acknowledged before
generation resumes. Prefix caches are invalidated, while the weight and
KV allocations stay resident. Resumed requests re-prefill their prefix
under the new version and retain their generated tokens, corresponding
sampling log-probabilities, and earliest behavior version, so Eq.~\ref{eq:staleness} also
covers responses that cross a refresh. The transfer runs in the
background while leases keep the source image valid, so refresh
overlaps the next eligible learner update without assembling a full
policy on a GPU or reallocating rollout state.

\subsection{Balancing the two engines}
\label{sec:design-coexec}
\label{sec:design-config}

\paragraph{Sharing knobs.} Configuration tunes two knobs before
training: the rollout layout (tensor or pipeline degree) and the
learner's MPS compute share $c$. Memory feasibility sets two more, the
rollout memory budget $u$ and the learner micro-batch budget $m$.
Reference and actor passes share the learner's cap.
Client priority and decode concurrency are not separate knobs: priority
showed no measurable effect and stays fixed, and decode concurrency
follows from the rollout memory budget $u$ and the batch size $N$.

\paragraph{What sets the step time.} In steady state every step trains
one batch and dispatches one cohort. Let $L$ be the mean learner-phase
duration (reference scoring, the actor update, and the CPU update) measured
with rollout running beside it, and let $S$ be the mean span of a
cohort from its dispatch to its last response. For $T\ge1$, the
publication gate ensures that a cohort dispatched at step $p$ finishes
before dispatch at step $p+T$. Averaging over steady-state cohorts gives
$S\le TP$, where $P$ is the mean step time. The learner also requires
$P\ge L$, so
\begin{equation}
  P\ge\max\!\left(L,\;S/T\right).
  \label{eq:period}
\end{equation}
At $T=0$ the engines run in turn and $P \ge L+S$.
The larger term guides the direction of the first compute-share probe.
Learner work can overlap the rollout tail because the gate waits after
training. Any tail that outlasts the learner update delays publication,
even when $L>S/T$ (\S\ref{sec:eval-config}).

Increasing the learner share $c$ gives training more compute while
increasing its competition with rollout. The rollout memory budget $u$
sets KV capacity, and the micro-batch budget $m$ sets the learner's
efficiency within the remaining memory. Configuration uses these
responses to choose which share to measure next.

\paragraph{Memory feasibility.} At a fixed layout, the per-GPU KV budget
in bytes follows $K(u)=K(u_0)+(u-u_0)M_{\mathrm{dev}}$. A feasible
configuration fits rollout weights, KV, learner window, activations,
buffers, and runtime within $M_{\mathrm{dev}}$, and provides enough KV
for a maximum-length sequence. The initial probe then checks the
runtime footprint before tuning the compute share.

\paragraph{Applying the balance to configuration.} The rollout layout
follows the interconnect: tensor parallelism where peer links are fast,
and pipeline stages or replicas where traffic crosses the host.
At a feasible memory budget, a first short run measures $L$ and $S/T$.
The next probe raises the learner share when $L$ is larger and lowers
it otherwise. Subsequent probes fit a power-law response for $L$ and
a linear response for $S/T$ using only the shares already visited,
and estimate neighboring shares from the nearest measured point.
Predicted reductions in $\max(L,S/T)$ guide which neighbor to measure.
Our local search uses ten-percentage-point share increments and at most
four measurements, including the initial run. A measured share replaces
the current choice only if it improves throughput by more than 5\%.
If no unmeasured neighbor is predicted to improve by more than 5\%,
the search stops unless the last probe exceeded that threshold and
there is an unmeasured neighbor and budget for a confirmation probe.

\section{Evaluation}
\label{sec:evaluation}

\subsection{Implementation and setup}
\label{sec:eval-setup}
\label{sec:implementation}

\paragraph{Implementation.} \sys\ integrates a streamed learner backend
with verl's asynchronous trainer~\citep{hybridflow} and vLLM
rollout~\citep{vllm}. Ray~\citep{ray} places the processes; trainer
hooks implement ordered consumption, the publication gate, and
background refresh from pinned host staging, and a verl extension
keeps rollout weight and KV allocations across publications.
In the evaluated configurations, \sys\ uses tensor-parallel
rollout on H100 and pipeline-parallel or single-GPU rollout
replicas on A800 and RTX~4090. On H100, the baselines
and all rollout engines use FlashAttention-3~\citep{flashattention3}.
For actor updates, \sys\ and the baselines share verl's output-layer
function and clipped policy objective, using recorded rollout
log-probabilities as the old-policy probabilities.

\paragraph{Platforms and workloads.} Unless stated otherwise, each experiment uses four GPUs:
H100 (80\,GB, NVLink), A800 (80\,GB, PCIe), and RTX~4090 (24\,GB, PCIe).
Each host has two CPU sockets: 96 Xeon Platinum 8468 cores and 2\,TiB
of DRAM (H100), 64 Xeon Platinum 8358P cores and 2\,TiB (A800), and 128
EPYC 7T83 cores and 1\,TiB (RTX~4090). The three platforms cover the
deployments co-residency targets: H100 nodes with fast GPU links,
large-memory A800 PCIe deployments, and small-memory RTX~4090 PCIe
nodes. Table~\ref{tab:workloads} characterizes
the workloads: Qwen3 models in Base, thinking, and non-thinking modes on
DAPO-Math~\citep{dapo}, with verifiable math rewards, and
MBPP~\citep{mbpp}, with unit-test pass@1 rewards. The running example is
Qwen3-8B-Base on RTX~4090 at $N{=}256$.

\paragraph{Baselines.} All systems use verl's training framework~\citep{hybridflow}
and vLLM rollout~\citep{vllm}; the baselines use verl's FSDP2 learner
tuned for each platform's memory budget, with optimizer
state on the host where GPU memory allows and full host offload
otherwise (Table~\ref{tab:baselines}), and \sys\ the host-owned learner of
\S\ref{sec:design-window}. \emph{Asynchronous time-sharing}
(\texttt{colocate\_async}) gives both phases all four GPUs in turn and
carries unfinished responses across steps through partial rollout.
\emph{Disaggregation} (2+2) assigns two GPUs to the learner and two
to one TP2 rollout replica. In the main asynchronous comparisons, we
configure both baselines and \sys\ with a staleness limit of
$T=\StalenessBound$ and the \texttt{wait} policy, with two batches in
flight. \emph{Synchronous time-sharing} is the on-policy reference
($T=0$). The time-sharing baselines transfer updated weights from
learner GPU tensors to rollout
and use the best measured feasible rollout layout for each workload. We
also compare execution layouts with \sys's learner and training recipe
held fixed under the same staleness setting: \emph{time-sharing}, where
rollout pauses during the learner phase, and \emph{whole-GPU splits} of the four GPUs
between the engines. For each workload and layout, we select the
best-performing feasible configuration among those measured, tuning
rollout parallelism, memory budgets, and the learner's compute cap where
applicable. Figure~\ref{fig:eval-ledger} adds 3L+1R for
non-thinking and 1L+3R for thinking. We select the end-to-end
configurations from platform- and workload-specific measurements;
\S\ref{sec:eval-config} evaluates the balance model and its
short-measurement procedure.

\begin{table}[t]
  \caption{Tuned verl time-sharing configurations.}
  \label{tab:baselines}
  \centering\footnotesize
  \setlength{\tabcolsep}{3.5pt}
  \newsavebox{\baselineconfigbox}
  \sbox{\baselineconfigbox}{\begin{tabular}{@{}lll@{}}
    \toprule
    Platform / workload & Training state & Rollout \\
    \midrule
    H100, 8K responses & \BlHState & TP4 \\
    H100, 16K responses & \BlHThinkState & TP4 \\
    A800 & \BlAState & TP2$\times$2 \\
    RTX~4090 & \BlRState & TP2$\times$2 \\
    \bottomrule
  \end{tabular}}
  \begin{minipage}{\wd\baselineconfigbox}
  \noindent\usebox{\baselineconfigbox}\par\vspace{3pt}
  {\scriptsize\raggedright A800 async uses TP1$\times$4 for
  thinking DAPO-Math at $N{=}256$.\par}
  \end{minipage}
\end{table}

\paragraph{Metric.} Throughput is consumed prompt and response tokens
per second over steps 5--12, a continuous window that ends with the
final weight publication. We average repeats within each seed, then
average over seeds; Figure~\ref{fig:eval-throughput} distinguishes single-seed
probes from three-seed means. We verify that every \sys\ run stays within
its staleness bound on consumed versions and drops no samples.

\subsection{End-to-end throughput}
\label{sec:eval-e2e}
\label{sec:eval-scaling}

\begin{figure*}[t]
  \centering
  \includegraphics[width=\textwidth]{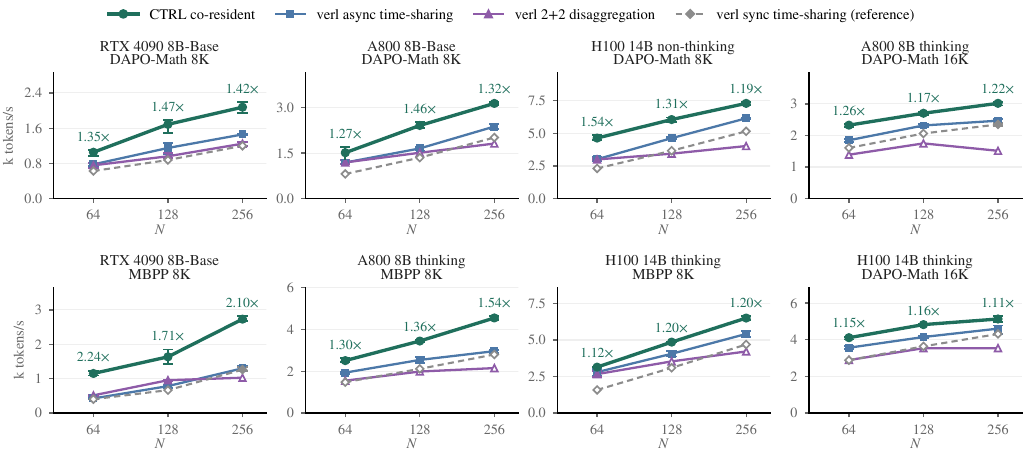}
  \caption{End-to-end throughput. Labels: \sys's speedup over the faster
  of verl's two asynchronous baselines. Filled markers: three-seed means;
  hollow markers: single seeds. Bars show the range of seed means.}
  \label{fig:eval-throughput}
\end{figure*}

\paragraph{\sys\ leads the asynchronous baselines.}
Figure~\ref{fig:eval-throughput} reports all workloads. \sys\ exceeds
both verl asynchronous baselines on every one of them, at
\HeadBestLo--\HeadBestHi$\times$ the throughput of the stronger
asynchronous baseline. This measures the complete system. The controls
below compare layouts with the learner held fixed. On A800 with
thinking MBPP at $N{=}256$ it reaches
\fm{amt256}{ctrl} tokens/s, \fmb{amt256}$\times$ asynchronous
time-sharing. Neither baseline dominates the other: asynchronous
time-sharing is stronger on most workloads, while 2+2 disaggregation
leads on A800 Base at $N{=}64$ and on RTX~4090 MBPP at $N{=}64$ and
$128$, where switching phases on every GPU costs more than halving each
engine.

\paragraph{Across batch sizes and workloads.} The gain holds from
$N{=}64$ to $256$ on all three platforms. A small batch shortens every
phase, so the per-step switching that \sys\ removes weighs more, while a
large batch gives the learner more work to overlap: on RTX~4090 the gain
is \fmb{r8b64}$\times$, \fmb{r8b128}$\times$, and \fmb{r8b256}$\times$ at
$N{=}64$, $128$, and $256$. Short-response Base workloads gain
\HeadBaseLo--\HeadBaseHi$\times$ and long-context thinking workloads
\HeadThinkLo--\HeadThinkHi$\times$. On long-response DAPO thinking
workloads, switching is a small share of each step, and the rollout
span sets the step time (\S\ref{sec:eval-config}). Short steps
expose time-sharing's fixed per-step costs most clearly. Turning thinking off on the same H100 model
and prompts raises the gain from \fmb{h16k256}--\fmb{h16k128}$\times$ to
\fmb{h8n256}--\fmb{h8n64}$\times$. Moving the same Base model on
RTX~4090 from DAPO-Math to MBPP, the shortest responses of all
workloads, raises it from \fmb{r8b64}--\fmb{r8b128}$\times$ to
\fmb{rmb128}--\fmb{rmb64}$\times$ (\S\ref{sec:eval-time}). Thinking MBPP
gains \fmb{amt64}--\fmb{amt256}$\times$ on A800 and
\fmb{hmb64}--\fmb{hmb256}$\times$ on H100.

\paragraph{Eight GPUs.} On eight H100 GPUs, each GPU holds a learner
rank and a rollout shard. At $N{=}512$, \sys\
reaches \EightCtrlTok\ tokens/s, \EightVerlX$\times$ asynchronous
time-sharing (three-seed means). With the same learner at
$N{=}512$, co-residency outperforms both splits in all three
seeds, reaching 1.10$\times$ the mean throughput of 4+4 and
\EightSixSplitX$\times$ that of 6+2. Its gain over 4+4 ranges from
1.08--1.13$\times$ across seeds.

\subsection{Where the gains come from}
\label{sec:eval-time}

\begin{figure*}[t]
  \centering
  \includegraphics[width=\textwidth]{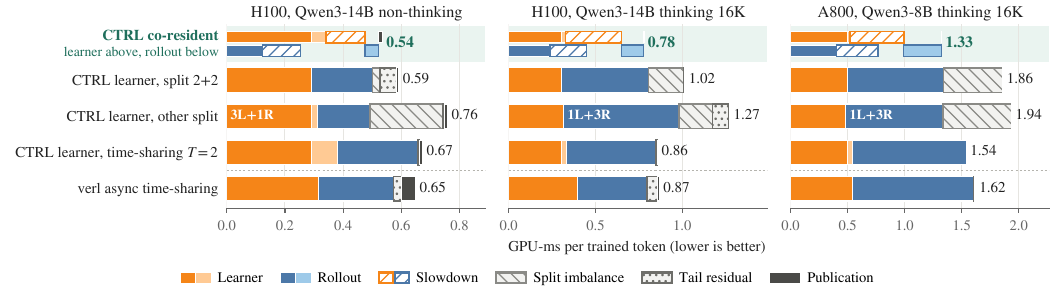}
  \caption{GPU allocation time by layout. In co-resident rows, rollout
  appears below the learner and overlap counts once. Rollout work is
  normalized by standalone replay throughput. Light blue marks exposed
  rollout, including its tail. Blue hatching shows the slowdown share
  estimated from matched decode intervals.}
  \label{fig:eval-ledger}
\end{figure*}

\begin{table}[t]
  \caption{Execution layouts with the same learner and training recipe
  (tokens/s). Each layout uses its best measured feasible configuration.
  Parentheses give co-resident throughput divided by
  the corresponding alternative's throughput.}
  \label{tab:layouts}
  \centering\footnotesize
  \setlength{\tabcolsep}{3.5pt}
  \begin{tabular}{@{}lrrrr@{}}
    \toprule
    Workload & $N$ & Co-resident & 2+2 split & Time-sharing \\
    \midrule
    \multicolumn{5}{@{}l}{\textit{DAPO-Math: three-seed means}} \\
    H100 14B-NT & 264 & \textbf{\SLNTCo} & \SLNTSplit\ (\SLNTSplitX$\times$) & \SLNTTs\ (\SLNTTsX$\times$) \\
    H100 14B-T & 256 & \textbf{\SLTKCo} & \SLTKSplit\ (\SLTKSplitX$\times$) & \SLTKTs\ (\SLTKTsX$\times$) \\
    A800 8B-T & 256 & \textbf{\SLATCo} & \SLATSplit\ (\SLATSplitX$\times$) & \SLATTs\ (\SLATTsX$\times$) \\
    A800 8B-B & 256 & \SLABCo & \textbf{\SLABSplit}\ (\SLABSplitX$\times$) & \SLABTs\ (\SLABTsX$\times$) \\
    4090 8B-B & 256 & \textbf{2176} & 2156\ (1.01$\times$) & 1724\ (1.26$\times$) \\
    \midrule
    \multicolumn{5}{@{}l}{\textit{MBPP: matched seed 29}} \\
    4090 8B-B & 64 & 1228 & \textbf{1421}\ (0.86$\times$) & 856\ (1.43$\times$) \\
    4090 8B-B & 256 & \textbf{2703} & 2432\ (1.11$\times$) & 2010\ (1.34$\times$) \\
    \bottomrule
  \end{tabular}
\end{table}

\paragraph{Same learner, different layouts.} Table~\ref{tab:layouts}
compares three layouts using the same learner implementation and
training recipe on each hardware platform. On DAPO-Math, co-residency is
\SLTsLo--\SLTsHi$\times$ faster than
time-sharing on all three platforms:
the learner phase no longer stops rollout. The matched-seed MBPP
controls show \SLRMLTsX--\SLRMTsX$\times$ over time-sharing; at
$N{=}256$, co-residency also reaches 1.11$\times$ the 2+2 split.
For the four-GPU H100 DAPO-Math and A800 thinking workloads, it is
\SLSplitLo--\SLSplitHi$\times$ faster than
the best measured whole-GPU split: every GPU serves both engines, so neither
waits for the other's pool. The advantage holds as the batch grows: on
H100 non-thinking it is \SLNTSplitX$\times$ at $N{=}264$ and
\SLNLargeX$\times$ at $N{=}528$, where the learner phase sets the step
time and a whole-GPU split can only add learner
GPUs by taking them from rollout. The controls also locate the regime
where a split already serves both engines efficiently: Base DAPO-Math
is within 1\% of the split on A800 and RTX~4090, and short MBPP at
$N{=}64$ favors the split. At $N{=}256$, co-residency overtakes it.
The GPU-time ledger explains these outcomes through overlap, sharing
costs, and pool imbalance.

\paragraph{Where GPU time goes.} Figure~\ref{fig:eval-ledger} accounts
for GPU allocation time per trained token: four GPUs times continuous
wall time divided by consumed tokens.
Splits pay for
imbalance: with 2+2 on H100 non-thinking the rollout GPUs wait while the
learner GPUs finish, and with three learner GPUs the learner waits for
one rollout GPU. For thinking, 1L+3R leaves one learner rank
on the critical path; assigning a second GPU to the learner improves
throughput on both H100 and A800.
Time-sharing adds the two engines' costs. In \sys, the learner phase
overlaps \LedNTHidden\,GPU-ms of rollout work per token while incurring a
sharing slowdown of \LedNTSlow. The resulting cost is
\LedNTCo\,GPU-ms per token, compared with \LedNTSplit\ for the best split
and \LedNTTs\ for time-sharing. Long thinking responses amplify pool imbalance: \LedTKCo\,GPU-ms per token for
co-residency, \LedTKSplit\ for the best split, and \LedTKTs\ for time-sharing.

Recorded phases supply learner and publication
costs; reference rates and request traces estimate rollout work and
waiting. For $k$ rollout GPUs, generating $D$ tokens at reference rate
$R$ accounts for $kD/R$ GPU-seconds, evaluated inside and outside the
learner phase to separate overlapped and exposed work. For \sys\ layouts with two or more learner ranks, the two-GPU
split supplies the learner-work base; subsequent segments show the
increase with more ranks and the sharing slowdown relative to standalone
learner execution. The 1L+3R and verl rows use their own measured learner
work. Request activity attributes split waiting; batch tails account for
the remaining allocation.

\paragraph{Sharing SMs is the mechanism.} The gain comes from pairing
unlike engines (Figure~\ref{fig:corun-counters}): the learner's dense computation fills the compute capacity that decoding leaves idle. The tail of
a cohort, where few requests remain, decodes beside the learner at
\KeepHRollSmall\ of its standalone rate on H100. In a single-GPU
microbenchmark on RTX~4090, a rollout engine beside a learner delivers
\KeepRSumLo--\KeepRSumHi\ times the work of either alone, whereas two
rollout engines, which contend for the same memory bandwidth, deliver
\KeepRPairLo--\KeepRPairHi. The gain also needs concurrent execution: with MPS off on RTX~4090,
both engines remain resident but run in turn, and throughput falls to
\AbMpsOffX$\times$, close to time-sharing's \AbTsX$\times$.

\begin{figure}[t]
  \centering
  \includegraphics[width=\columnwidth]{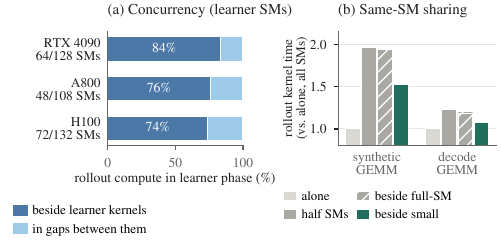}
  \caption{MPS sharing: (a) rollout-kernel overlap during the learner
  phase; (b) same-SM sharing on RTX~4090.}
  \label{fig:mps-sharing}
\end{figure}

\paragraph{How MPS shares each GPU\@.} Kernel traces of the paper
configurations show concurrent execution and use of gaps between learner
kernels (Figure~\ref{fig:mps-sharing}a). The learner sees 64/128 SMs
on RTX~4090, 48/108 on A800, and 72/132 on H100. During its phase,
74--84\% of rollout compute time overlaps learner kernels, with the
rest in their gaps.
A separate controlled microbenchmark demonstrates same-SM co-residency
with a persistent learner stand-in at a 50\% share
(Figure~\ref{fig:mps-sharing}b). Beside the small-footprint stand-in,
rollout kernels recover 44--60\% of the time gap between execution
beside a full-SM blocker and standalone execution on the full GPU.
Sharing has a price, and the share sets it. The learner runs on part of
the SMs and contends with rollout for memory bandwidth, L2, and power,
which is the sharing slowdown of Figure~\ref{fig:eval-ledger}. Without
a share, its millisecond-scale kernels hold every SM: in single-GPU
kernel pairs, rollout then keeps as little as 18\% of its standalone
rate, against at least 61\% at a 60\% share.

\begin{table}[t]
  \caption{Co-residency raises GPU activity in the running example
  (Figure~\ref{fig:overview-modes}). Nsight Systems metrics are
  time-weighted over steps 5--11 and averaged across four GPUs.}
  \label{tab:gpu-activity}
  \centering\footnotesize
  \setlength{\tabcolsep}{3pt}
  \begin{tabular}{@{}lrrr@{}}
    \toprule
    System & \shortstack{SM\\Active (\%)} & \shortstack{SM\\Issue (\%)} & \shortstack{Tensor\\Active (\%)} \\
    \midrule
    verl sync time-sharing & 47.5 & 3.4 & 8.5 \\
    verl async time-sharing & \ProfileColoActive & \ProfileColoIssue & \ProfileColoTensor \\
    verl 2+2 disaggregation & 41.9 & 2.8 & 8.8 \\
    \sys\ & \textbf{\ProfileCtrlActive} & \textbf{\ProfileCtrlIssue} & \textbf{\ProfileCtrlTensor} \\
    \bottomrule
  \end{tabular}
\end{table}

\paragraph{GPU activity in the running example.}
Table~\ref{tab:gpu-activity} profiles each layout on the running example
with Nsight Systems over steady-state steps. \sys\ keeps SMs active \ProfileCtrlActive\% of the time,
against 41.9--47.5\% for the baselines, and raises SM issue from
\ProfileColoIssue\% to \ProfileCtrlIssue\% and tensor-pipe activity from
\ProfileColoTensor\% to \ProfileCtrlTensor\% over asynchronous
time-sharing: co-residency keeps the GPUs busier.

\begin{table}[t]
  \caption{Throughput and realized age at each configured staleness
  limit $T$: H100 14B-NT, DAPO-8K, $N{=}256$.
  Entries give tokens/s (mean / maximum age).}
  \label{tab:staleness}
  \centering\footnotesize
  \setlength{\tabcolsep}{4pt}
  \begin{tabular}{@{}lrrr@{}}
    \toprule
    $T$ & 0 & 2 & 4 \\
    \midrule
    \sys\ & \StZeroCtrl\ (0/0) & \StTwoCtrl\ (\StTwoCtrlReal/2) & \StFourCtrl\ (\StFourCtrlReal/4) \\
    verl async time-sharing & -- & \StTwoVerl\ (\StTwoVerlReal/2) & \StFourVerl\ (\StFourVerlReal/5) \\
    verl sync time-sharing & \StZeroVerl\ (0/0) & -- & -- \\
    \bottomrule
  \end{tabular}
\end{table}

\paragraph{Each version bound.} Serialized at $T=0$,
\sys\ runs its phases in turn with rollout resident. On H100
non-thinking it then matches tuned synchronous time-sharing,
\StZeroCtrl\ against \StZeroVerl\ tokens/s, and publishes in
\NTZeroSwitchS\,s instead of a \NTSyncSwitchS\,s weight switch. Table~\ref{tab:staleness}
compares the systems at each configured limit: at $T=\StalenessBound$ and $T=4$,
\sys\ runs the two engines concurrently and leads asynchronous
time-sharing at both configured limits. Realized age uses each consumed
trajectory's earliest behavior version (Eq.~\ref{eq:staleness}). \sys\ enforces both bounds at
consumption; native's wait check covers in-flight requests, allowing
completed trajectories in its queue to reach age~5 at configured $T=4$.
At $T=4$, \sys\ doubles realized age and gains \StFourOverTwoPct\%
throughput; learner time nearly sets step time at $T=2$
(\S\ref{sec:eval-config}).

\paragraph{Short steps on 24\,GB GPUs.} The largest gains,
\MbppLo--\MbppHi$\times$ on RTX~4090 MBPP, come from a per-step cost that
the baselines pay regardless of batch size. On 24\,GB GPUs their training
state must live in FSDP2's CPU offload (Figure~\ref{fig:memory-budget}),
and every step moves model-sized parameters and gradients across GPUs
without NVLink: the actor phase stays at \MbppActorLo--\MbppActorHi\,s and
the weight switch at about \MbppSwitchS\,s while the batch grows
fourfold from $N{=}64$ to $256$. With 70-token responses, these two costs
make up most of the baseline's \MbppStepLo--\MbppStepHi\,s step. \sys's
learner reads layers from host memory and resident rollout weights while
rollout runs, and publication takes about \MbppCtrlPubS\,s, so \sys\
completes a step in \MbppCtrlLo--\MbppCtrlHi\,s.

\begin{table}[t]
  \caption{Removing one mechanism at a time (tokens/s): RTX~4090
  Qwen3-8B-Base, $N{=}256$.}
  \label{tab:ablation}
  \centering\footnotesize
  \setlength{\tabcolsep}{4pt}
  \begin{tabular}{@{}lrr@{}}
    \toprule
    Variant & tokens/s & vs.\ \sys \\
    \midrule
    \sys & \AbRCtrlTok & 1.00$\times$ \\
    \quad no publication gate & \AbRNoGateTok & \AbRNoGateX$\times$ \\
    \quad no weight reuse & \AbRNoReuseTok & \AbRNoReuseX$\times$ \\
    \bottomrule
  \end{tabular}
\end{table}

\paragraph{Mechanism ablations.} Table~\ref{tab:ablation} removes one
mechanism at a time from the tuned configuration. Without the
publication gate, the learner waits before sampling, as in verl's
dropless sampler, and exposes the younger cohort's tail ahead of every
learner update (Figure~\ref{fig:staleness}a); throughput falls to
\AbRNoGateX$\times$ on the RTX~4090. Without rollout-weight
reuse, the learner window loads every layer from the host, which lowers
it to \AbRNoReuseX$\times$ over PCIe.

\subsection{Resource responses and configuration}
\label{sec:eval-config}

\begin{table}[t]
  \caption{Step time $P$, learner phase $L$, and rollout span $S/T$ at
  the best measured share (seconds, $T{=}2$).}
  \label{tab:span}
  \centering\footnotesize
  \setlength{\tabcolsep}{4pt}
  \begin{tabular}{@{}lrrr@{}}
    \toprule
    Workload & $P$ & $L$ & $S/T$ \\
    \midrule
\SpanRows
    \bottomrule
  \end{tabular}
\end{table}

\paragraph{The balance predicts useful compute shares.}
Table~\ref{tab:span} reports the learner phase, cohort span, and
measured step time at the selected shares. Across three platforms,
we validate the balance with retrospective ranking on \ModelPickAll\
same-seed share sweeps. Ranking candidates by their measured tokens per
update divided by $\max(L,S/T)$ selects a share within 2\% of the
measured best in \ModelPickOk\ sweeps, with a mean throughput
shortfall of \ModelPickLoss\%.
Gate wait accounts for
\fmg{r8b256}--\fmg{r8b64}\% of the step on RTX~4090,
\fmg{a8t128}--\fmg{a8b64}\% on A800, \fmg{hmb256}--\fmg{hmb64}\% on H100
MBPP, and \fmg{h16k128}--\fmg{h16k64}\% on H100 thinking. These waits
show how much of the rollout tail remains exposed.

\paragraph{Short measurements apply the balance.}
On two new single-seed workloads, two measurements per workload select
shares within 1.5\% and 4.0\% of the best subsequently measured on the
same seed on RTX~4090 and A800, respectively. The procedure starts at a
50\% learner share after establishing memory feasibility; each
measurement runs 12 updates and scores updates 5--12. Both retain the
initial share under the 5\% selection threshold.

\begin{figure}[t]
  \centering
  \includegraphics[width=\columnwidth]{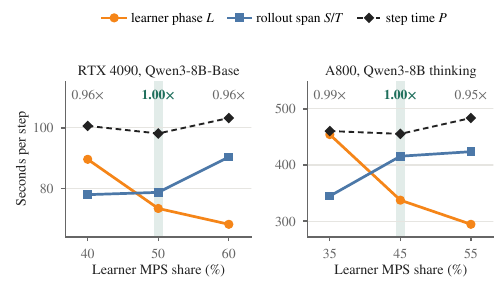}
  \caption{Learner MPS-share sweeps: three-seed means on RTX~4090
  and a single seed on A800. Top labels give throughput relative to
  each panel's best measured share.}
  \label{fig:eval-share}
\end{figure}

\paragraph{The share moves the two terms in opposite directions.}
Figure~\ref{fig:eval-share} sweeps the learner's share on RTX~4090
Base and A800 thinking. On both, a larger share shortens the learner
phase and lengthens the rollout span, the step time follows the larger
of the two, and throughput peaks next to the share where they cross. These responses guide which neighboring shares the local search probes.

\paragraph{Measured resource responses.} The two engines respond
differently to the learner's share, which is what makes the balance
reachable. The learner gives up SMs at less than proportional cost:
alone it does not fill them, so on H100 a 70\% share runs only
\HwLearnShareCost\% slower than 99\%. On RTX~4090
(Figure~\ref{fig:eval-share}), lowering the share from 60\% to 50\%
lengthens the learner phase by \SweepRLearnPct\% and shortens the rollout
span by \SweepRSpanPct\%. The memory law fixes KV
capacity before a run: from \LawKVHAnchor\,GiB at a 0.70 rollout
memory budget on H100 it predicts \LawKVH\,GiB at 0.74, the value
measured, and it marks an RTX~4090 budget infeasible
(\LawKVInfeasible\,GiB, below one sequence's \LawKVNeeded\,GiB),
which fails at startup.

\begin{figure}[t]
  \centering
  \includegraphics[width=\columnwidth]{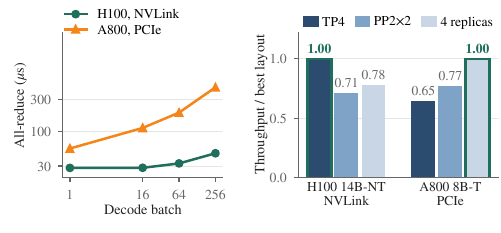}
  \caption{Interconnect and rollout layout. Left: one all-reduce at
  decode sizes. Right: co-resident throughput of three rollout layouts,
  relative to the best on each platform.}
  \label{fig:eval-layout}
\end{figure}

\paragraph{Interconnect-aware rollout layouts.} Tensor parallelism
adds an all-reduce to every decoding step. Over NVLink it stays near
30--50\,$\mu$s at decode sizes; when GPUs communicate through the host it
grows with the batch to several hundred microseconds
(Figure~\ref{fig:eval-layout}), the same order as the decode step
itself. The best co-resident layout follows: on H100, TP4 leads and
gives long responses the shortest per-token latency; on A800, four
single-GPU replicas lead and TP4 trails. These measurements guide layout before share tuning.

\subsection{Training behavior and host memory}
\label{sec:eval-staleness}
\label{sec:eval-quality}
\label{sec:eval-costs}

\paragraph{Training quality.} We train Qwen3-14B non-thinking on four
H100s for \EvQualUpdates\ updates at $N{=}128$ and evaluate greedily on
\OlyQuestions\ English, text-only competition math problems from
OlympiadBench~\citep{olympiadbench}. Averaging the last three evaluations
within each of two seeds, \sys\ reaches \OlyCtrlAcc\% accuracy against
\OlyColoAcc\% for verl asynchronous time-sharing under the same
staleness bound, with \OlyColoTokX$\times$ its consumed-token throughput
over the full training runs. On RTX~4090, Qwen3-8B-Base trained for
\EvQualUpdates\ steps reaches \IntroQualityCtrl\% under \sys\ and
\IntroQualityAsync\% under disaggregation with the same bound on a
decontaminated \EvQualQuestions-question MATH~\citep{math} subset.
No run evicts samples.

\paragraph{Host memory.} \sys\ keeps training state in host
memory. For Qwen3-14B, master weights, Adam moments, and FP32 gradient
accumulators take \IntroStateGiB\,GiB, and each leased BF16 policy image
\HostImageGiB\,GiB, well within server DRAM. Host streaming overlaps
GPU execution: in a steady-state H100 update at $N{=}256$, the
learner phase takes 56.7\,s, of which the GPU passes take 52.1\,s while
weights stream beneath them at 5.1\,GB/s per GPU, 8\% of the PCIe link,
and CPU clipping and Adam over all 14.8B parameters add 3.8\,s.

\section{Related Work}
\label{sec:related}

\paragraph{RL execution layouts.} HybridFlow (verl), OpenRLHF, and
DeepSpeed-Chat~\citep{hybridflow,openrlhf,deepspeed_chat} support
colocated execution in which phases take turns; ReaL~\citep{real}
searches stage execution plans, and DistFlow~\citep{distflow} and
slime~\citep{slime} develop distributed control and training--inference
integration. These systems time-share or split the GPUs; \sys\ adds a
co-resident layout in which a learner rank and a rollout shard run
concurrently on every GPU.

\paragraph{Asynchronous training and scheduling.}
AReaL~\citep{areal} decouples training from interruptible generation;
StreamRL, ROLL Flash, Laminar, and
StaleFlow~\citep{streamrl,rollflash,laminar,staleflow} develop
generation, versioning, and scheduling; LlamaRL,
AsyncFlow, and PipelineRL~\citep{llamarl,asyncflow,pipelinerl} address
weight synchronization and streamed execution; and Seer~\citep{seer}
accelerates synchronous rollout. \sys\ bounds asynchrony between co-resident engines, moving waits from
sampling to publication.

\paragraph{GPU sharing and phase overlap.} Rollplex~\citep{rollplex}
overlaps response-independent prefix work with decoding on the same GPUs
under CUDA MPS for vision-language post-training, keeping updates
synchronous and on-policy; \sys\ overlaps the complete learner update,
including passes over generated responses, with later rollout under the
bounded staleness of asynchronous RL. RLHFuse~\citep{rlhfuse} redistributes rollout tails
and OPPO~\citep{oppo} pipelines output chunks; DynaResize and
BiDiRL~\citep{dynaresize,bidirl} resize the allocation between phases,
RollMux~\citep{rollmux} shares resources across RL jobs, and
ROSE~\citep{rose} runs rollout on serving GPUs. MPS, MuxServe, AntMan,
and Orion~\citep{nvidia_mps,muxserve,antman,orion} multiplex independent
GPU work. MISO~\citep{miso} uses MIG to adapt GPU partitions to co-located
jobs. LithOS reallocates compute units at fine granularity~\citep{lithos}.
REEF and Tally~\citep{reef,tally} protect a latency-critical job from
best-effort work on the same GPU.
RESONATOR~\citep{resonator} shares SMs and HBM between the encoder and
decoder during multimodal serving. FlexLLM~\citep{flexllm} co-serves LLM
inference and parameter-efficient fine-tuning, and NanoFlow~\citep{nanoflow} overlaps
compute-, memory-, and network-bound operations within one serving
engine. \sys\ shares every GPU between the two engines of one
training job. Their progress is coupled through policy versions, so
neither can be treated as best-effort: the sharing is balanced, and
host-owned training state is what makes room for it.

\paragraph{Training-state residency.} ZeRO~\citep{zero} shards training
state, ZeRO-Offload and ZeRO-Infinity~\citep{zero_offload,zero_infinity}
move it to CPU or NVMe, and verl exposes FSDP2 offload as a training
backend~\citep{verl_fsdp_config}. Layer-streaming fine-tuning
systems~\citep{slideformer,slidedp} train dense models with
host-resident state and a sliding layer window. These systems use host
memory to fit one model; \sys\ uses it for a second, persistent engine on every GPU throughout
the RL policy lifecycle.

\section{Discussion}
\label{sec:implications}

\paragraph{Implications for GPU systems.} Co-residency changes what an
RL node asks of its hardware. Decoding and training load different
parts of a GPU (Figure~\ref{fig:corun-counters}), so a device that
serves one engine leaves computation that the other can use. First, the GPU becomes a shared
device inside one job. MPS lets the two engines use the SMs concurrently. The measured resource
responses in \S\ref{sec:eval-config} motivate runtimes that adjust
compute shares between phases; a single-process design using green
contexts offers another path to finer control over SM allocation. Second, training state moves to the host, so host memory capacity
and bandwidth and the host--GPU link, rather than the GPU footprint of
training state, bound the model that a node can train beside its own
rollout. Third, the interconnect
decides the rollout layout: tensor parallelism pays where GPUs have fast
peer links, and replicas or pipeline stages suit nodes whose GPUs
communicate through the host. The step-time bound guides measurements
of the learner's compute share. The GPU-time ledger accounts for
how overlap, sharing costs, and exposed waiting translate the
configuration into end-to-end throughput.

\section{Conclusion}
\label{sec:discussion}
\label{sec:conclusion}

\sys\ establishes co-residency as a third RL execution layout: it
organizes the RL policy lifecycle around host-owned versions so
that every GPU can host a learner rank beside a persistent rollout
shard. Scoring, training, and policy delivery fit a bounded GPU
footprint; publication scheduling overlaps complete learner updates
with later rollout under a dropless staleness bound; and a
step-time bound guides short measurements of their resource balance.
Across \HeadRows\ workload settings on H100, A800, and RTX~4090,
\sys\ achieves \HeadBestLo--\HeadBestHi$\times$ the throughput of the
stronger asynchronous baseline.
Every GPU holds the same pair of engines, so the layout avoids a fixed
whole-GPU partition.
The design turns the complementary
demands of rollout and learning into concurrent progress on the full
GPU allocation.

\bibliographystyle{ACM-Reference-Format}
\bibliography{references}
\end{document}